\documentclass{kapproc} % Computer Modern font calls
\usepackage{txfonts,graphicx,epsfig}
\kluwerbib

\startauthorindex

\def\sun{\hbox{$\odot$}}
\def\micron{\hbox{$\mu$m}}
\def\nat{Nature}%
\def\apj{ApJ}%
\def\aap{A\&A}%
\def\mnras{MNRAS}%
\begin{document}

%------------ article title  ------------------->>
\articletitle{Radiative Transfer in Prestellar Cores:
a Monte Carlo approach}

\author{D.~Stamatellos}
\affil{Department of Physics \& Astronomy, Cardiff University, Wales, UK}
\email{D.Stamatellos@astro.cf.ac.uk}
\author{A.~P.~Whitworth}
\affil{Department of Physics \& Astronomy, Cardiff University, Wales, UK}
\email{A.Whitworth@astro.cf.ac.uk}

%% optional, to supply a shorter version of the title for the running head:
\chaptitlerunninghead{Radiative Transfer in Prestellar Cores: A Monte Carlo Approach}

\anxx{Author1\, and Author2}

\begin{abstract}
We  use our Monte Carlo radiative transfer code
to study non-embedded prestellar cores and 
cores that are embedded at the centre of a molecular cloud. 
Our study indicates that the temperature inside embedded cores
is lower than in isolated non-embedded cores, and generally less than
12 K, even when the cores are surrounded by an ambient cloud of small visual
extinction ($A_{\rm V}\sim 5$).
Our study shows that the best wavelength region to observe embedded
cores is between 400 and 500 $\micron$, where the core is quite distinct from the
background.
We also predict that very sensitive observations 
($\sim 1-3$ MJy${\rm\, sr^{-1}}$)
at 170-200 $\micron$ can be used to estimate how deeply a core is
embedded in its parent molecular cloud.  Finally, we present preliminary
results of asymmetric models of non-embedded cores.

\end{abstract}

%------------ body of article ------------------->>

\section*{Introduction}

Prestellar cores are cores that are either on the verge of collapse 
or already collapsing
(e.g. Myers \& Benson 1983, Ward-Thompson et al. 2002).
They represent the initial stage of star formation and their study is 
important
since theoretical models of star formation are very sensitive to the 
initial conditions.

Prestellar cores have been observed either isolated or embedded in protoclusters.
Isolated prestellar cores (e.g. L1544, L43, L63) have extent 
$\stackrel{>}{_\sim}1.5\times 10^4$ AU and  masses
$0.5-35~{\rm M}_{\sun}$ (Ward-Thompson et al. 1999, Andr\'{e} et al. 2000).
On the other hand, prestellar cores embedded in protoclusters
(e.g.  in $\rho$ Oph, NGC2068/2071) are generally smaller,
with extent $\sim 2-4 \times 10^3$ AU and 
masses $\sim 0.05-3~{\rm M}_{\sun}$ 
(Motte et al. 1998, Motte et al. 2001). 

In this paper, we present radiative transfer models of non-embedded
and embedded
prestellar cores, performed using a 3-D Monte Carlo radiative
transfer code we have developed ({\sc Phaethon}).

\section{The Method: Monte Carlo Radiative Transfer}

Our method (Stamatellos \& Whitworth 2003)
is similar to that developed by Wolf, Henning \& Stecklum (1999) and
Bjorkman \& Wood (2001). We represent the radiation field of a source (star or 
background radiation) by a large number of monochromatic luminosity packets
($L$-packets). These $L$-packets are injected into the system
and interact stochastically with it. If an $L$-packet is absorbed its energy is
added to the local cell and raises the local temperature. To ensure
radiative equilibrium  the $L$-packet is
re-emitted immediately with a new frequency
chosen from the difference between the local cell emissivity before and 
after the absorption of the packet (Bjorkman \& Wood 2001). 
This method conserves energy exactly, accounts for the diffuse
 radiation field and its
3-dimensional nature makes it attractive for application in a variety of
systems.

The code has been thoroughly tested using the thermodynamic 
equilibrium test (Stamatellos \& Whitworth 2003) and also against
benchmark (Ivezic et al. 1997) and previous 
(Bjorkman \& Wood 2001) calculations.

\section{Non-Embedded Prestellar Cores}

We represent prestellar cores  by Bonnor-Ebert (BE) 
spheres, in which gravity is balanced by gas pressure (Bonnor 1956, Ebert 1955). 
In many cases this is a good approximation to prestellar cores
(e.g. Alves et al. 2001, Ward-Thompson et al. 2002).

For the radiation incident on the core, we use the Black (1994)
interstellar radiation field (hereafter BISRF), that
consists of an optical component due to radiation from giant stars and dwarfs,
 a component due to 
thermal emission from dust grains, mid-infrared radiation from non-thermally 
heated grains and the cosmic background radiation.

The opacity of the dust at the low temperatures (5-20K) and
high densities ($10^4-10^7$ cm$^{-3}$) expected in prestellar cores,
is quite uncertain. 
In our study, we use the opacities calculated by Ossenkopf
and Henning (1994), for grains that have coagulated 
and accreted thin ice mantles.

We find that the temperature inside  non-embedded cores drops from around 17 K,
at the edge of the core, to a minimum at the centre, which maybe as low
as 7 K, depending on the visual extinction to the centre of the core.
Our results are similar to those of previous studies 
(Evans et al. 2001, Zucconi et al. 2001).

\section{Embedded Prestellar Cores}
The radiation field incident on cores embedded in molecular clouds is
different from that incident on isolated non-embedded cores. The
ambient molecular cloud absorbs the UV, optical and NIR part of the
radiation and reemits it in the FIR and submm region (Mathis et al. 1983).
It also makes the radiation incident on the core anisotropic, because,
in general, the ambient cloud is not homogeneous. 
Another  factor that contributes to
the anisotropy of the radiation incident on an embedded core is the 
presence of stars or protostars in the vicinity of the core or the cloud
(e.g as in $\rho$ Oph; Liseau et al. 1999).

\paragraph{The Model}
In this first approach (Stamatellos \& Whitworth 2003), 
we study a spherical core at the centre of
a molecular cloud of uniform density.
The radiation incident on the molecular cloud is the BISRF, but 
the radiation incident on the core is enhanced in the FIR and submm and
reduced at shorter wavelengths, as a result of the presence of the 
molecular cloud around the core.

\begin{figure*} [ht]
\centerline{
\includegraphics[width=3.85cm]{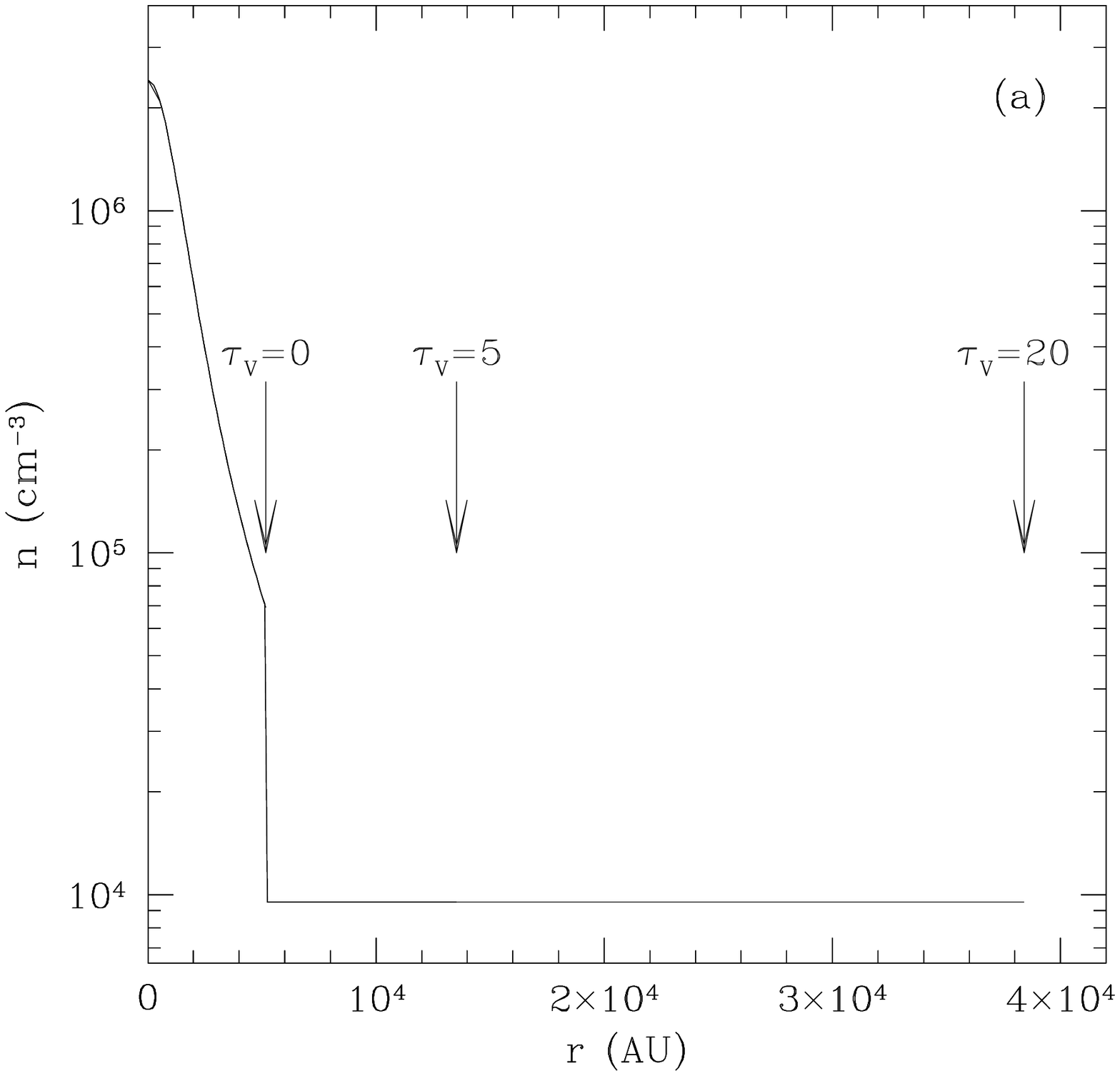}
\includegraphics[width=3.85cm]{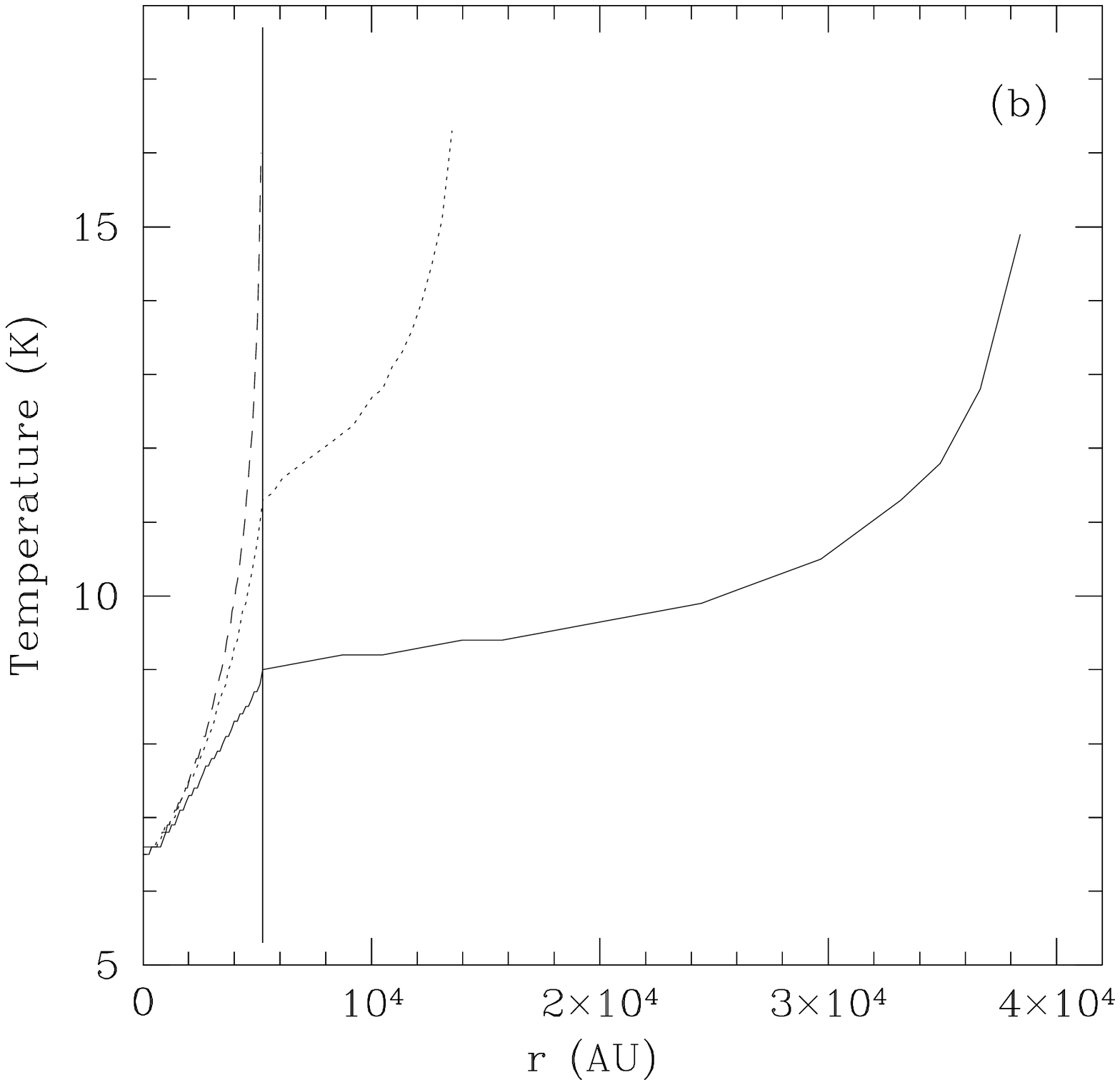}
\includegraphics[width=3.85cm]{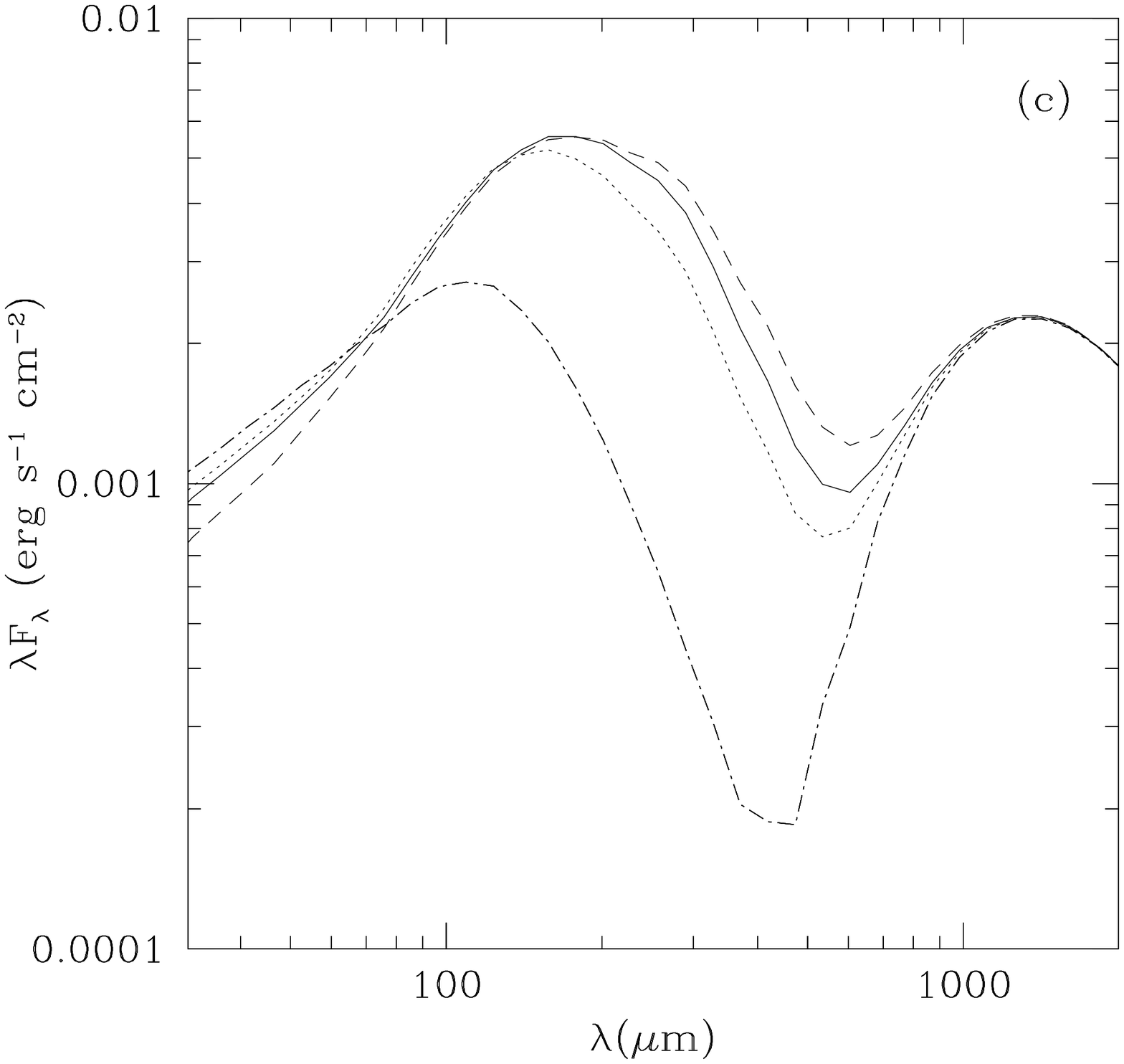}
}
\caption{(a) Density, (b) dust temperature, and (c) 
SEDs, for a supercritical core (see text)
surrounded by a spherical cloud with visual optical depth 20 (solid lines),
 5 (dotted lines) and 0 (dashed lines).
The dash-dot line on the SED graph corresponds to the background SED.}
\label{fig_oem2a}
\end{figure*}

\begin{figure*}[ht]
\centerline{
\includegraphics[width=3.85cm]{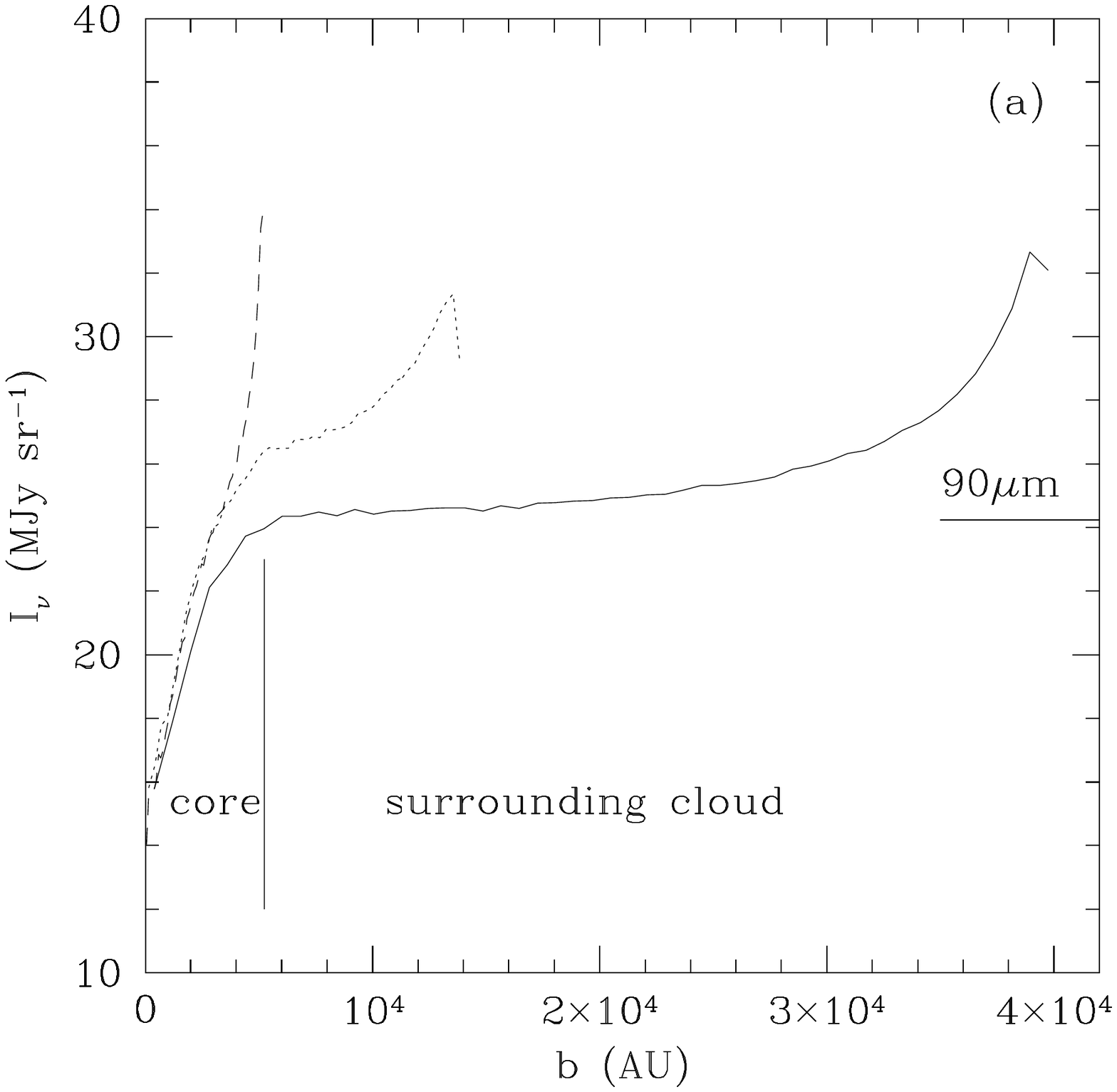}
\includegraphics[width=3.85cm]{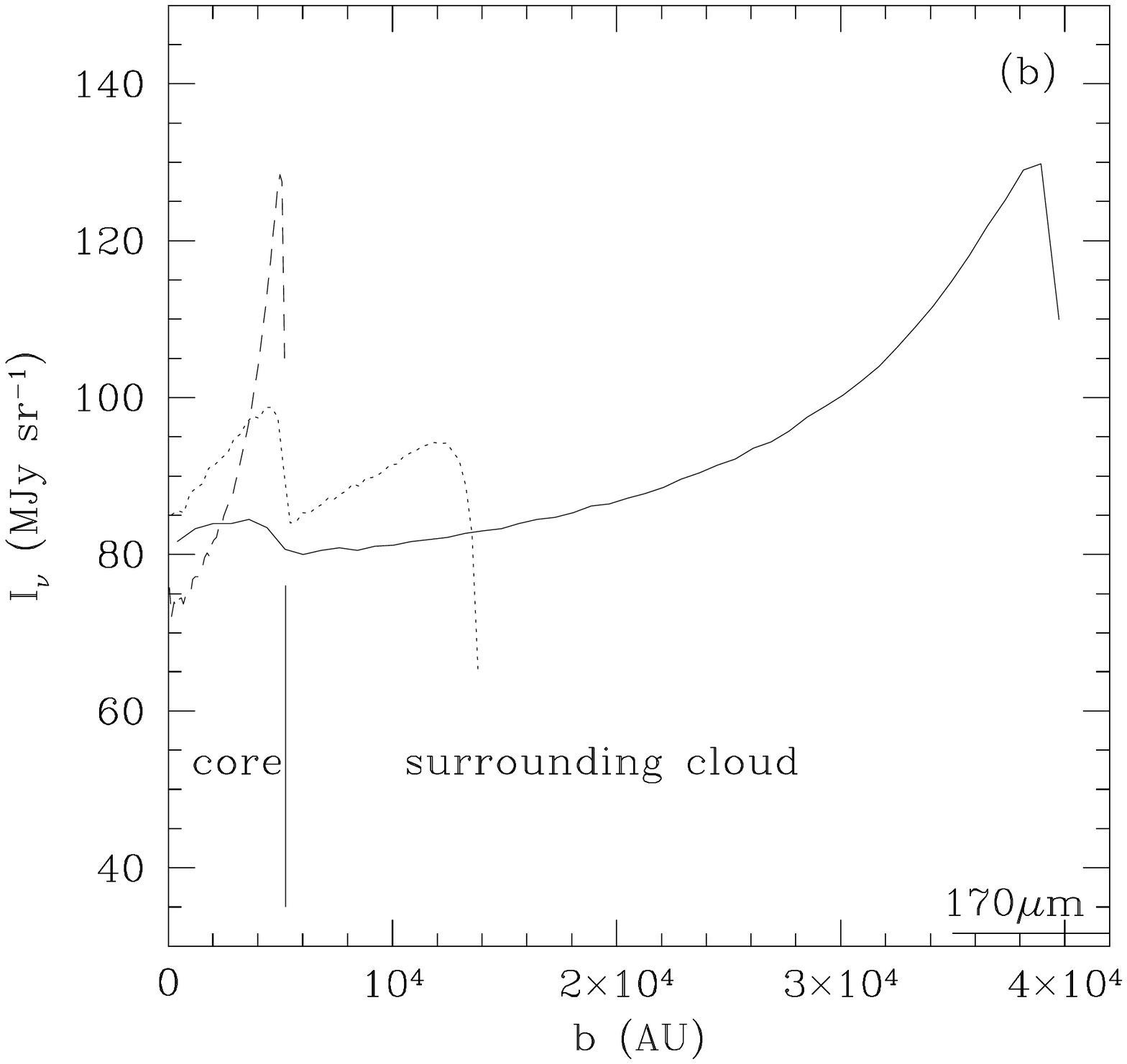}
\includegraphics[width=3.85cm]{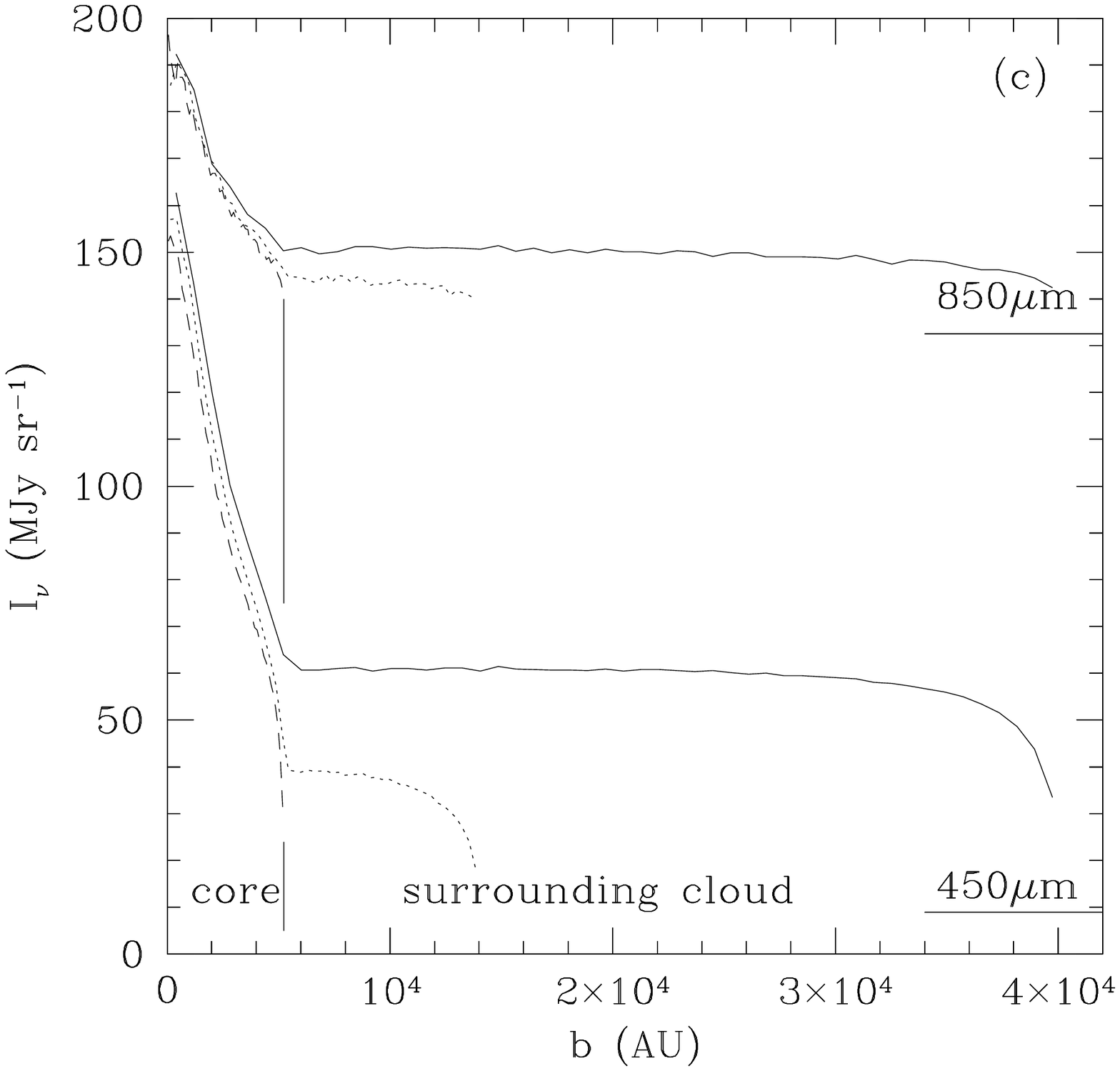}
}
\caption{Intensity profiles
at (a) 90, (b) 170, and (c) 450 and 850~$\micron$, for the models
in Fig.~\ref{fig_oem2a}. The horizontal solid lines on the 
profiles correspond to the background intensity at the
wavelength marked on the graph.}
\label{fig_oem2b}
\end{figure*}

We chose the parameters of our models so as to mimic 
the embedded prestellar cores and
the conditions in the $\rho$ Oph protocluster (see Motte et al. 1998):
core sizes $4-8 \times 10^3$ AU, masses $0.4-1.2~{\rm M}_{\sun}$,
ambient cloud particle density $n_{\rm tot}=0.96 \times 10^4\;{\rm cm}^{-3}$
and  ambient cloud pressure  $\sim 10^6\; $cm$^{-3}$~K. 
In Figs.~\ref{fig_oem2a}-\ref{fig_oem2b}, we present our calculations for 
a supercritical core at gas temperature $T$=15~K with mass 0.8 M$_{\sun}$,
under external pressure $P_{\rm ext}=10^6\;{\rm cm}^{-3}$~K, surrounded
by a spherical ambient cloud with different visual optical depths.

\paragraph{Core Temperature Profiles}
We find that the presence of even a moderately thick cloud 
($A_{\rm V}=5$) around the core, results in a less steep 
temperature profile inside the core than in the case of a core
that is directly exposed to the BISRF.
When there is no surrounding cloud (Fig.~\ref{fig_oem2a}, dashed lines), 
the temperature drops
from $\sim$16~K at the edge of the core to around 6-7~K in the centre 
($\Delta T\approx 9-10$~K, depending on the core density), whereas with a   
$\tau_V=5$ ambient cloud (Fig.~\ref{fig_oem2a}, dotted lines) 
the temperature drops from  around 11~K to 7~K 
($\Delta T\approx 4$~K).
Our studies show that dust temperatures inside embedded cores are
probably lower than 12~K in cores surrounded by even a relatively thin  cloud
($A_{\rm V}\approx 5$),
which seems to be the case for many of the prestellar cores in $\rho$ Ophiuchi.
Previous studies (Motte et al. 1998, Johnstone et al. 2000) 
of  cores in this region, 
assumed isothermal dust at temperatures from 12 to 20~K, 
when calculating core masses from mm observations and, thus, they may have
underestimated masses by a factor of 2.

Recent studies (Andr\'e et al. 2003, Andr\'e, this volume)
find similar temperature profiles, using a different
approach, in which they estimate
the effective radiation field incident on an embedded core from observations.

%%%%%%%%%%%%%%%%%%%%%%%%%%%%%%%%%%%%%%% fig_oem2a.ps
%%%%%%%%%%%%%%%%%%%%%%%%%%%%%%%%%%%%%%%%%%%%%

\paragraph{SEDs and Intensity Profiles}

At 90 microns the core is seen in absorption against the 
background (Fig.~\ref{fig_oem2b}a), and the intensity 
increases towards the edge of the core.
For very centrally-condensed cores
the decrease towards the centre
is just $\sim 8-10$ MJy${\rm\, sr^{-1}}$, but 
for less centrally-condensed cores it is even lower. Thus,
very sensitive 
(say $\sim 1-3 $ MJy${\rm\, sr^{-1}}$) observations are
needed to detect cores in absorption at 90~$\micron$. 

%%%%%%%%%%%%%%%%%%%%%%%%%%%%%%%%%%%%%%% fig_oem2b.ps
%%%%%%%%%%%%%%%%%%%%%%%%%%%%%%%%%%%%%%%%%%%%%%%

At wavelengths near the peak of the core emission (150-250~$\micron$)
the intensity increases by a small 
amount ($\sim 5-20$ MJy${\rm\, sr^{-1}}$ above the background) 
towards the edge of the core (Fig.~\ref{fig_oem2b}b). 
The higher the increase in the intensity near the core boundary, 
the less embedded 
is the core. Thus, very sensitive observations of embedded prestellar cores
at 170-200~$\micron$, might allow us to determine the extinction of the cloud
surrounding the core, and, thus, to estimate roughly the
position of the core inside the cloud. 
However, more sophisticated modelling is required, with 
an accurate density profile for the cloud and taking into account the
close environment of the core under study.

Finally, at submillimeter and millimetre wavelengths (400-1300~$\micron$)
the intensity drops toward the edge of the core considerably
(Fig.~\ref{fig_oem2b}c).
The core can be easily observed at 400-500~$\micron$, where the contrast
with the background is quite large
($\sim 50-150$ MJy${\rm\, sr^{-1}}$).
At wavelengths longer than $\sim$ 600~$\micron$ 
the background radiation becomes important and the core emission is
not much larger than the background emission 
($\sim$ 20-50 MJy${\rm\, sr^{-1}}$ larger).

\paragraph{Diagnostics}

In Table~\ref{tab:diagnostics}, we list  the peak intensities
at various wavelengths 
for cores embedded in molecular 
clouds with visual optical depths 5 and 20. 
This table indicates that embedded  cores are 
most easily distinguished from the
background radiation around 450~$\micron$.
%%%%%%%%%%%%%%%%%%%%%%%%%%%%%%%%%%%%%%%%%
\begin{table}[ht]
\caption{Typical peak$^{\mathrm{*}}$ intensities for embedded cores}
\begin{center}\begin{tabular}{@{}ccc}
\hline
 $\lambda$ ($\micron$) &  \multicolumn{2}{c}{ ${I_\lambda}^{\mathrm{a}}$ (MJy${\rm\, sr^{-1}}$)} \\
 &   $\tau_{\rm cloud}=5$ &  $\tau_{\rm cloud}=20$ \\
\hline\hline
90$^{\mathrm{b}}$   &   5-15  &  $\sim$ 3  \\  
170     &   10-15 &   $\sim$ 3  \\
450     &   55-160      &   40-130  \\ 
850     &   20-80       &   15-70    \\
1300    &   10-40 &   10-25  \\
\hline
\end{tabular}\end{center}
\begin{tablenotes}
$^{\mathrm{*}}$ The term {\it peak} refers to the maximum intensity
above or below the background (as noted)
at a specific wavelength.

$^{\mathrm{a}}$ Approximate peak intensities
for a core embedded in a cloud with visual optical depth $5$ and
 $20$.
The deeper the core is embedded the less distinct from
the background is. The lower value corresponds to a subcritical core and the higher
value to a supercritical  (i.e. more centrally condensed)  core.

$^{\mathrm{b}}$ At 90~$\micron$ the core seen in absorption against the 
background.
\end{tablenotes}
\label{tab:diagnostics}
\end{table}
%%%%%%%%%%%%%%%%%%%%%%%%%%%%%%%%%%%%%%%%%%%%%%%
The peak emission from 
embedded cores could be as low as $\sim$ 10 MJy${\rm\, sr^{-1}}$ above the background at
1300~$\micron$, but it's at least  $\sim$ 40 MJy${\rm\, sr^{-1}}$ at  450~$\micron$.
The wavelength range between 400-500 seems favourable for observing
embedded cores but the atmospheric transmission is not good in this range 
and space observations are needed. The upcoming
{\it Herschel}  (to be launched in 2007) will be operating in this range. 

\section{Asymmetric Models of Non-Embedded  Cores} %%%%%%% ASYMMETRIC CORES

\begin{figure*} [ht]
\centerline{
\includegraphics[width=3.85cm]{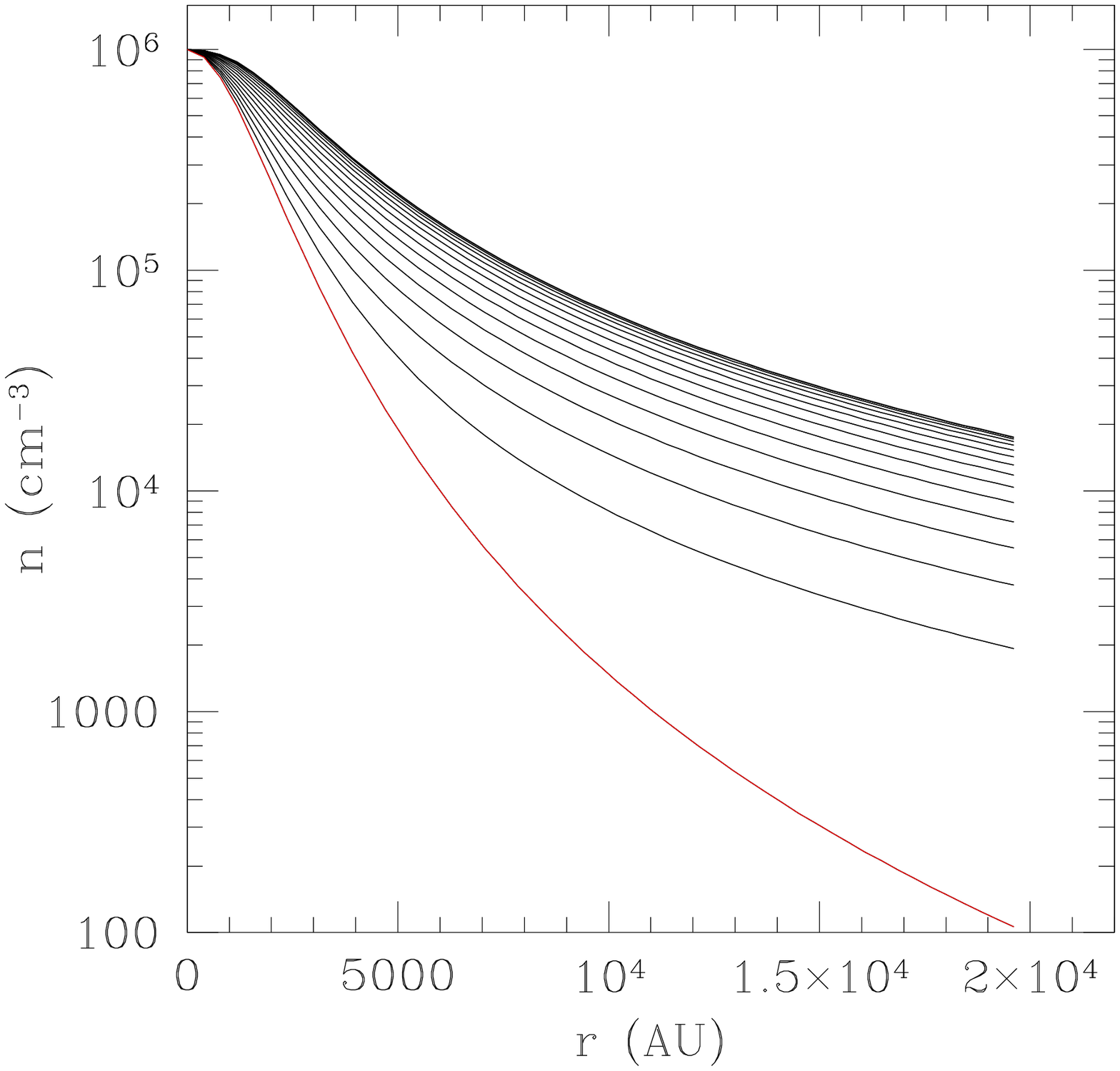}
\includegraphics[width=3.85cm]{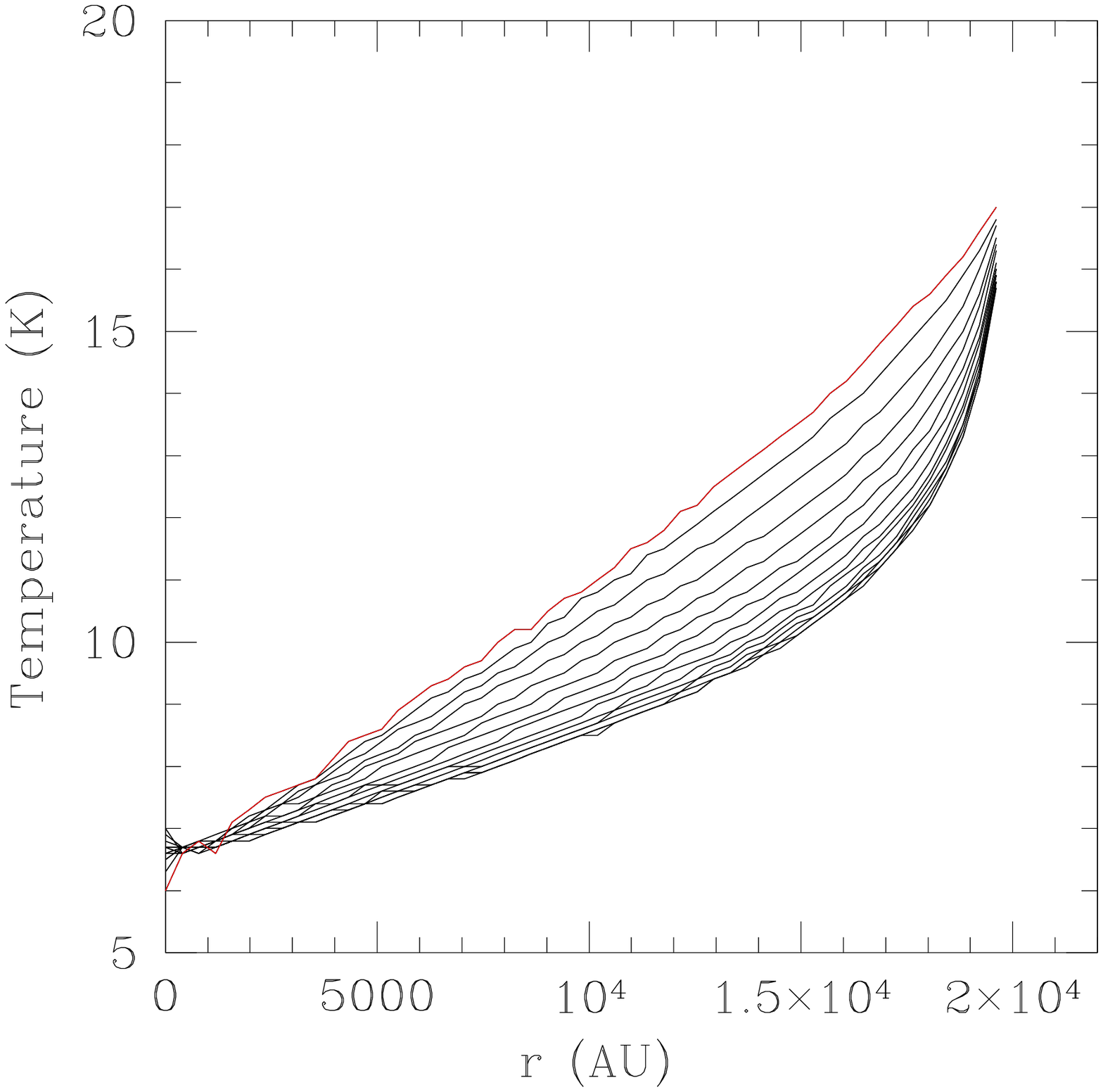}
\includegraphics[width=3.85cm]{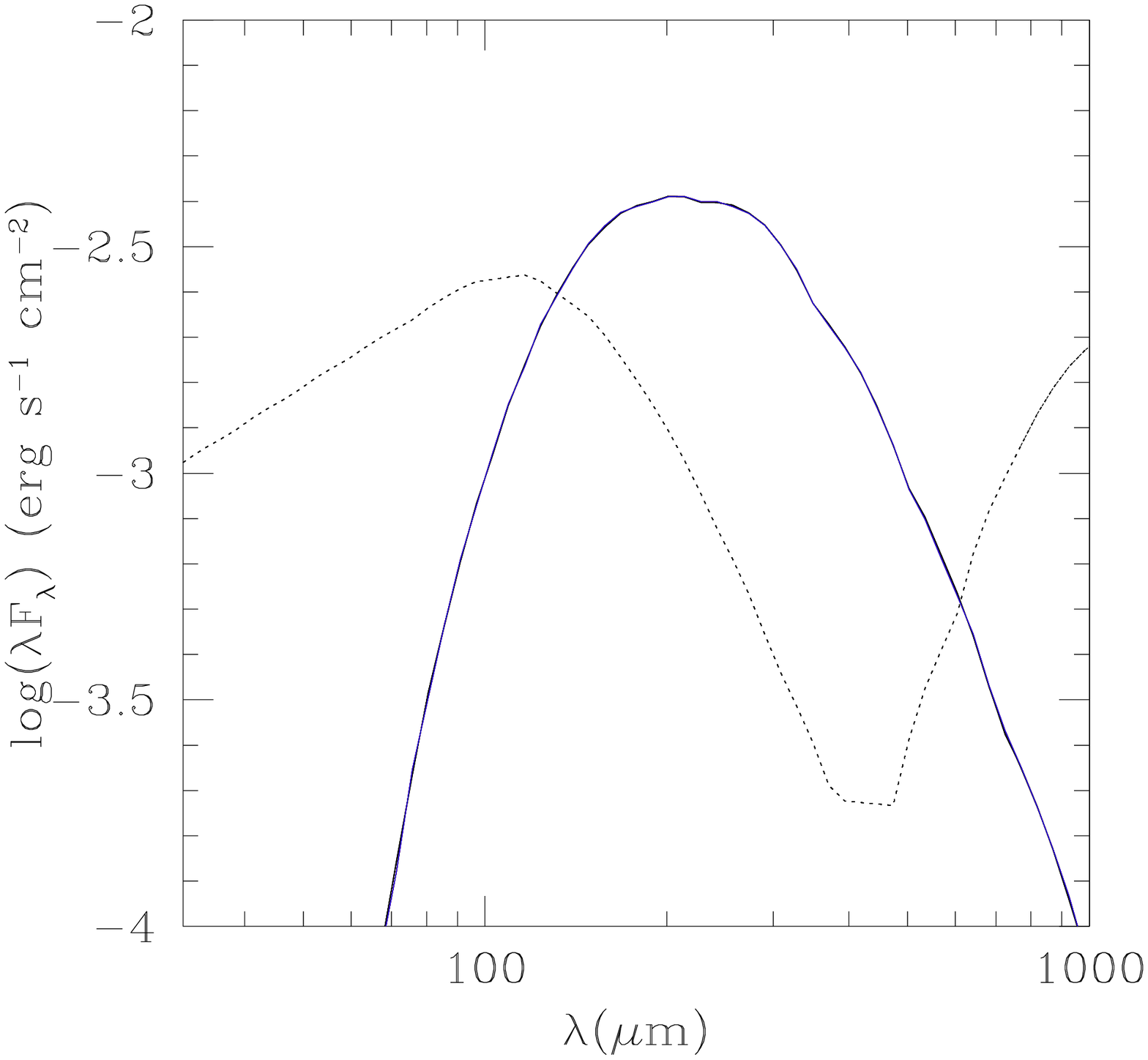}
}
\caption{(a) Density, (b) dust temperature, and (c) 
SEDs, for an asymmetric core (see text). The core is divided into 15
equal-angle cells (each cell's angle span is 12$^{\rm o}$). 
The lower density cell corresponding to $\theta=0^{\rm o}$ (lower
curve on the density plot), has higher temperature (upper curve in
the temperature plot).
The dotted line on the SED graph corresponds to the incident SED.}
\label{asym.cores}
\end{figure*}

\begin{figure*}[ht]
%
%\centerline{
%\includegraphics[width=5.4cm,angle=-90]{br1.0.ps}
%\includegraphics[width=5.4cm,angle=-90]{br2.0.ps}}
%\centerline{
%\includegraphics[width=5.4cm,angle=-90]{br1.30.ps}
%\includegraphics[width=5.4cm,angle=-90]{br2.30.ps}}
%\centerline{
%\includegraphics[width=5.4cm,angle=-90]{br1.90.ps}
%\includegraphics[width=5.4cm,angle=-90]{br2.90.ps}}
%
\centerline{
\includegraphics[width=5.4cm,angle=-90]{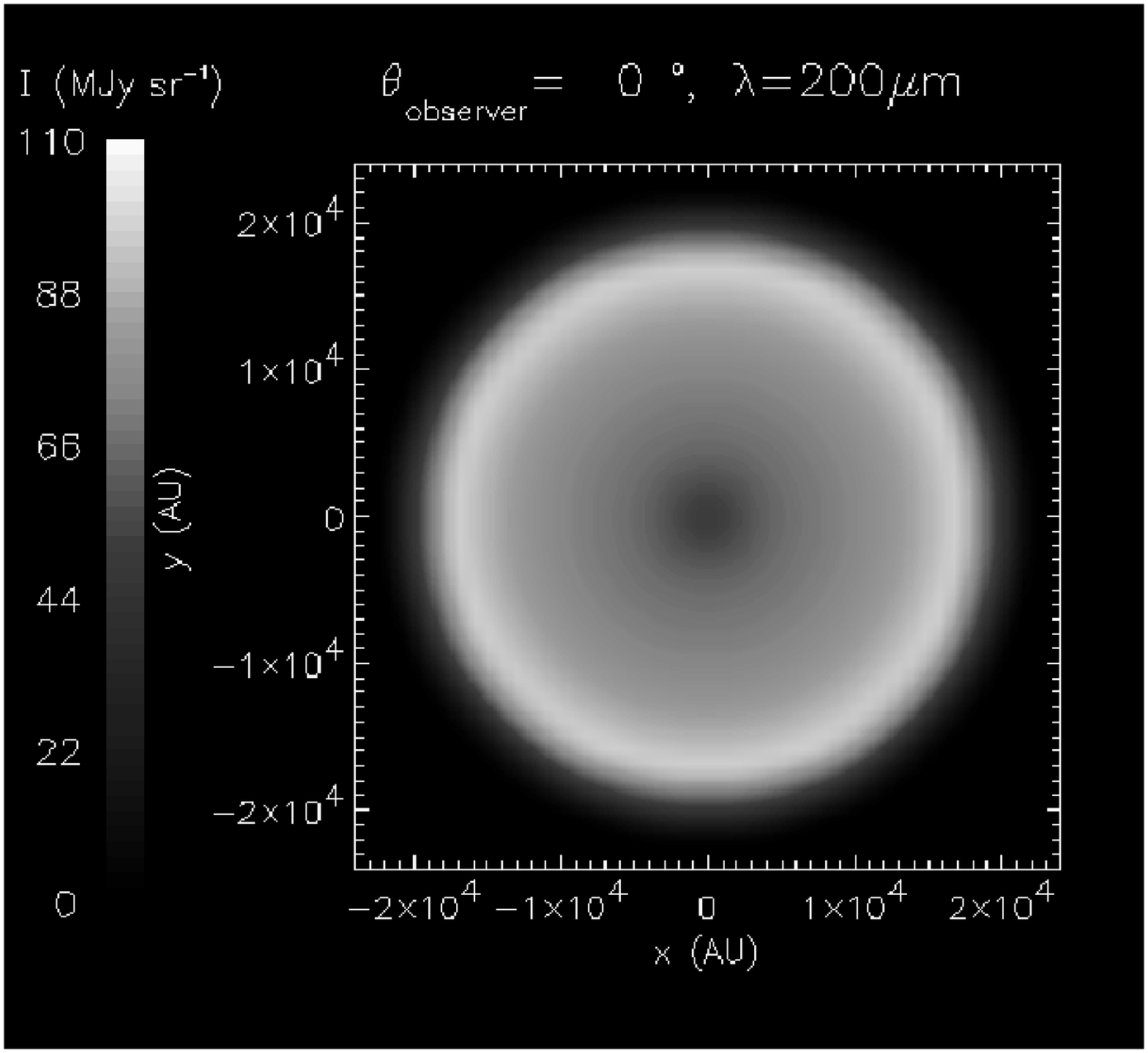}
\includegraphics[width=5.4cm,angle=-90]{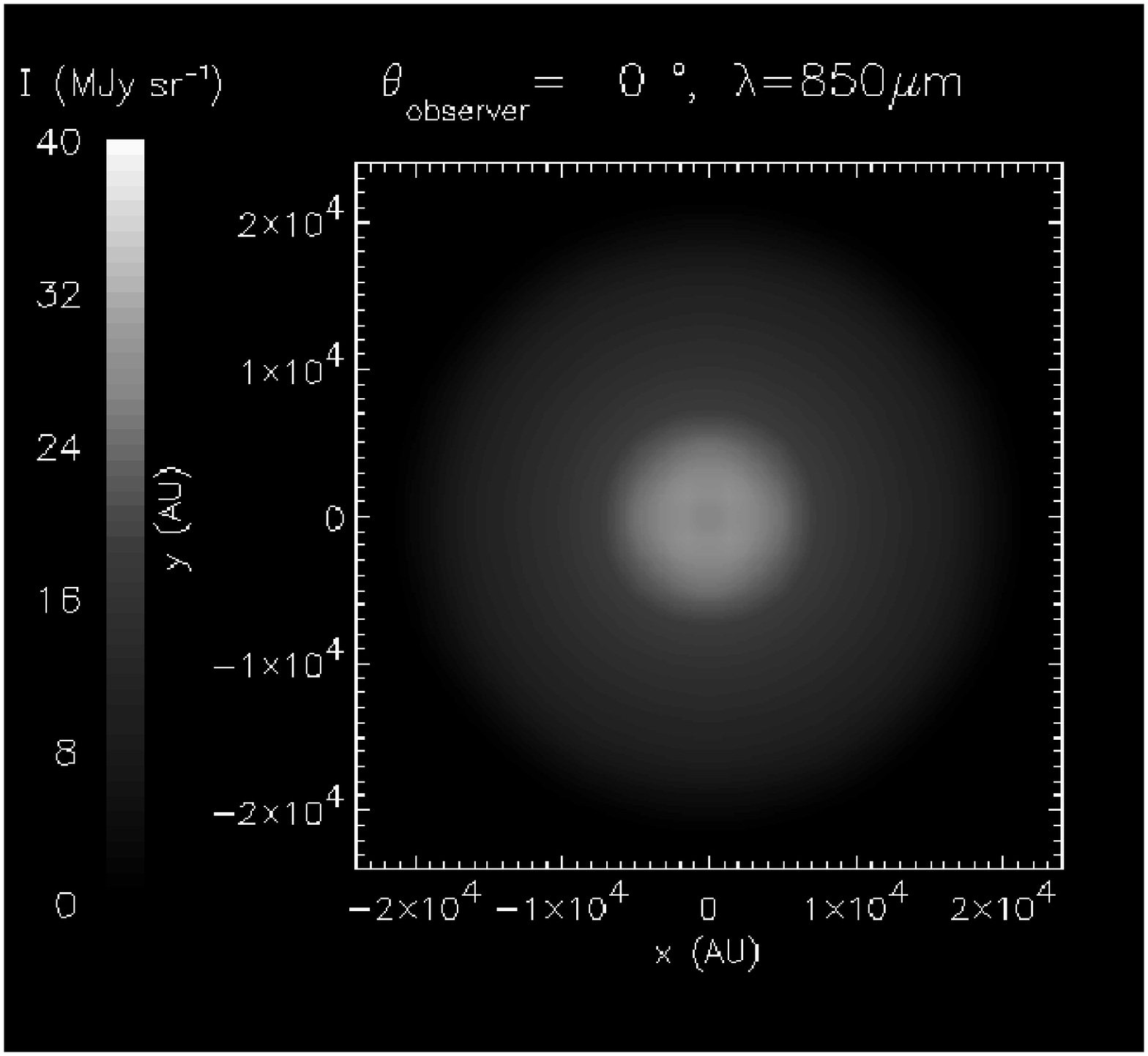}}
\centerline{
\includegraphics[width=5.4cm,angle=-90]{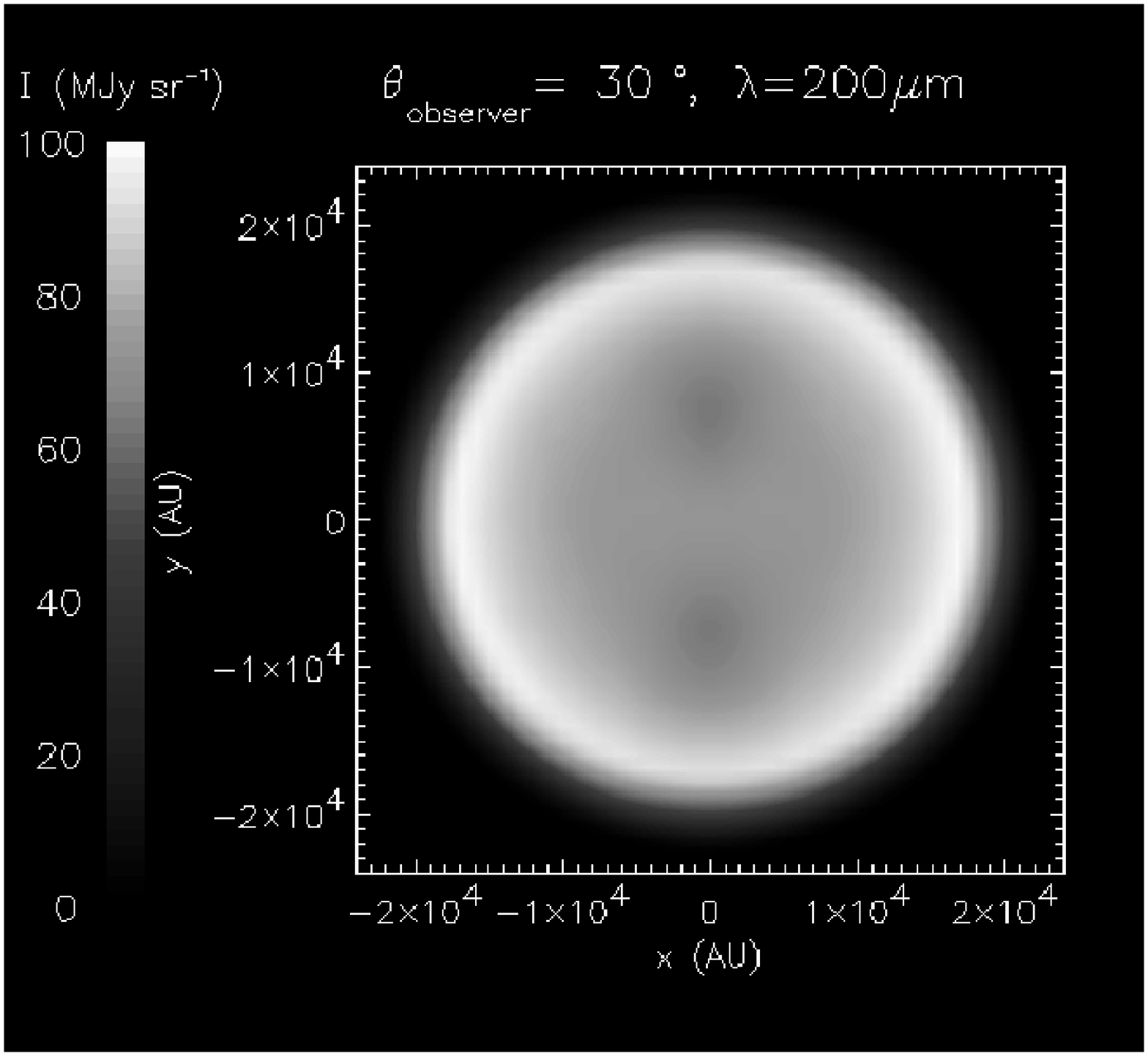}
\includegraphics[width=5.4cm,angle=-90]{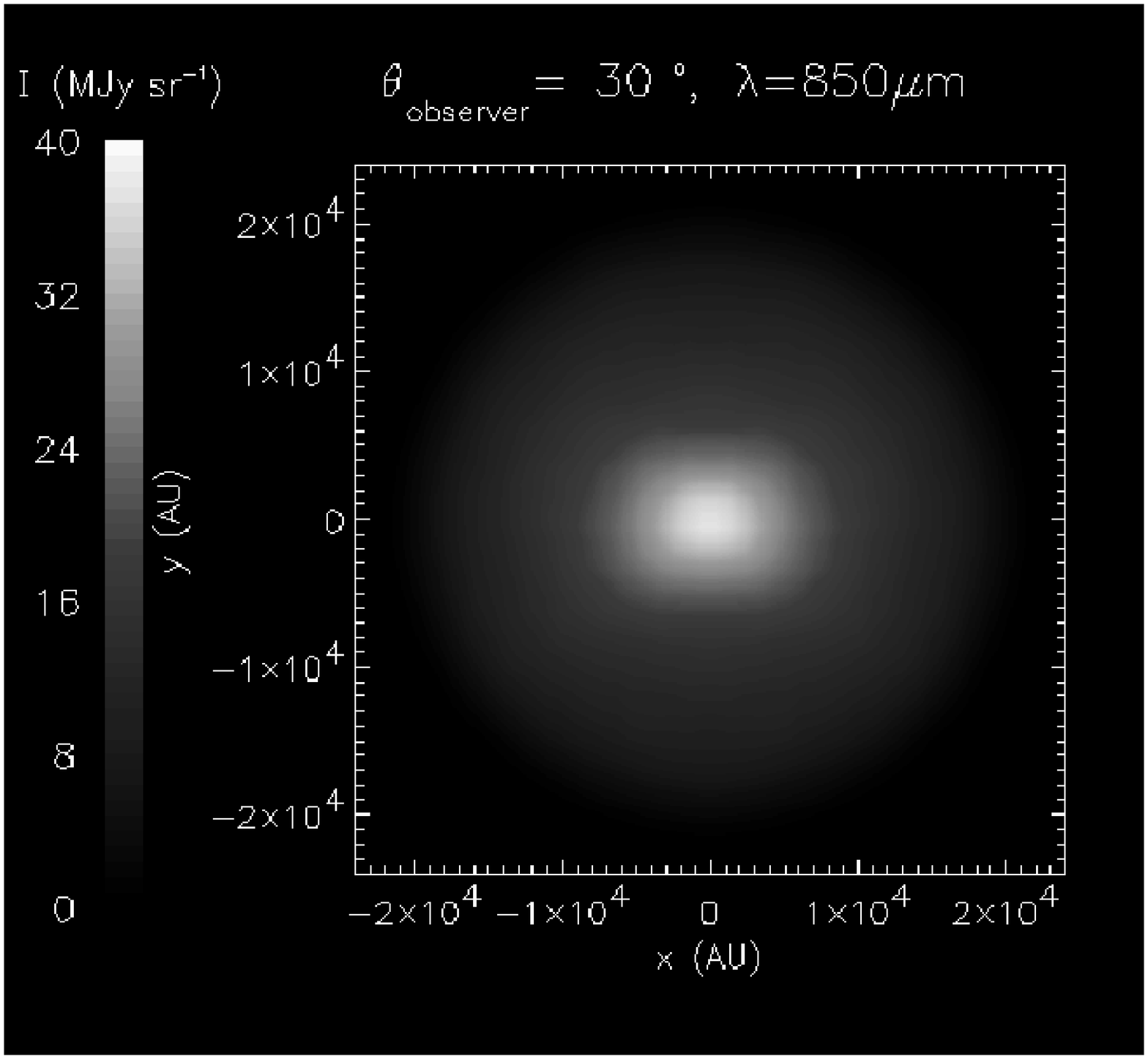}}
\centerline{
\includegraphics[width=5.4cm,angle=-90]{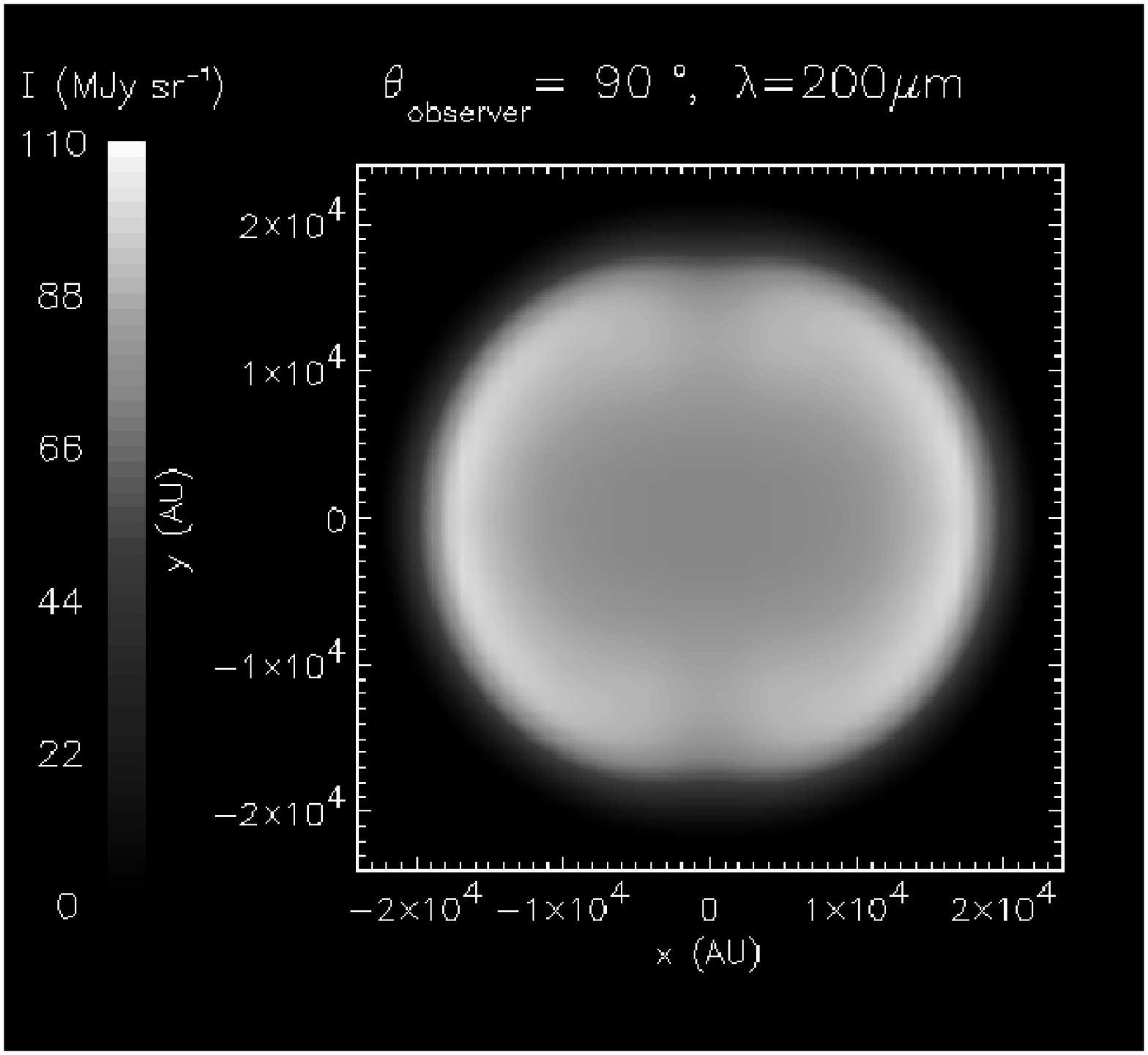}
\includegraphics[width=5.4cm,angle=-90]{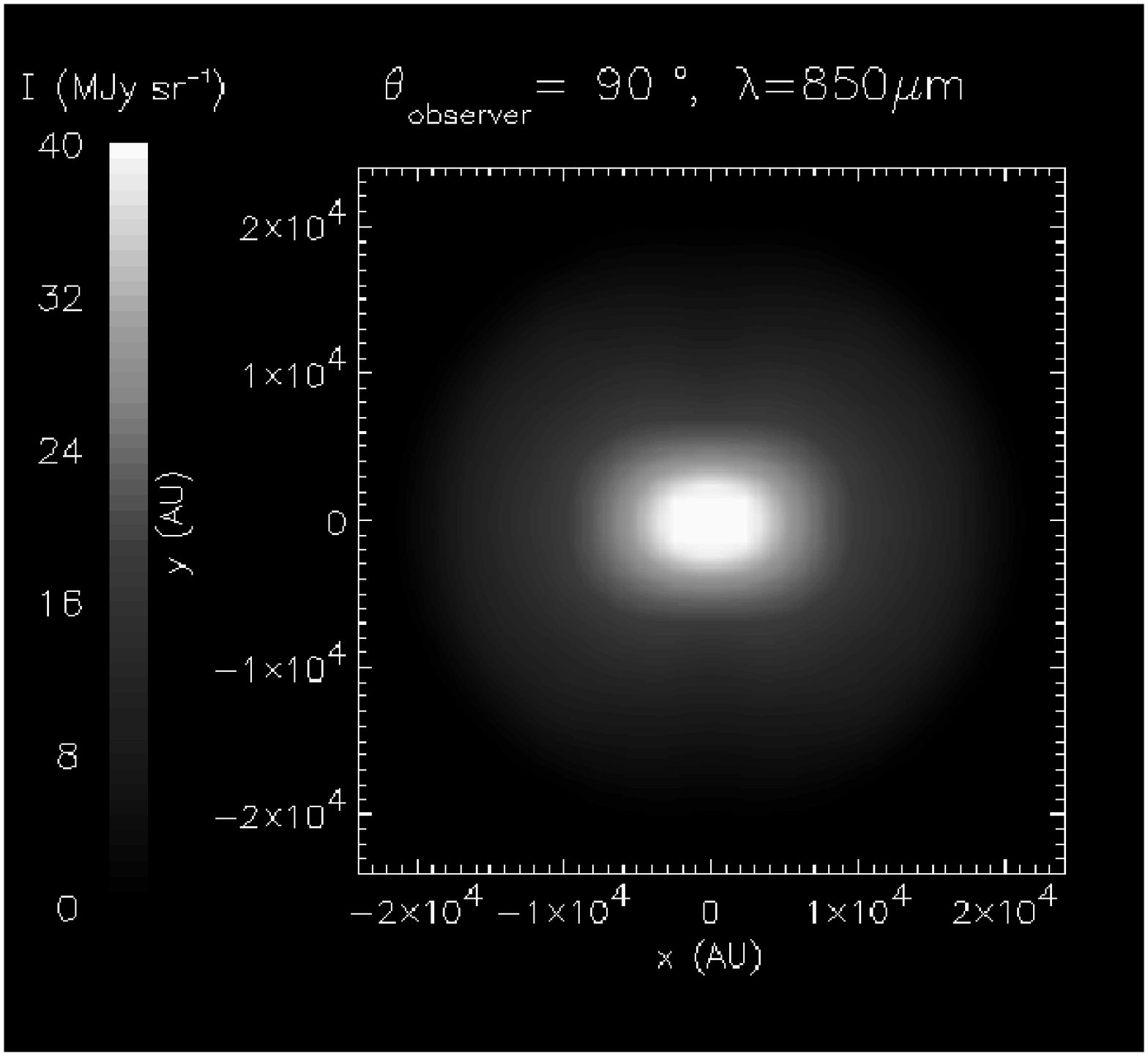}}
\caption{Isophotal maps at 200 (left panel) and 850 $\micron$ (right
panel) at viewing angles $0^{\rm o}$, $30^{\rm o}$ and $90^{\rm o}$ 
(top to bottom), for a disk-like asymmetric
core (see text).  The core appears elongated when viewed at a direction
different from $\theta=0^{\rm o}$. 
The colour image
is available on the CD-ROM.}
\label{images}
\end{figure*}

\paragraph{The model} 
Generally, cores are not spherically symmetric.
Here, we assume non-embedded cores with a disk-like
asymmetry, i.e. the core is denser on the equatorial plane;
\begin{equation}
\rho(r,\theta)=\rho_c\frac{1+A\left(\frac{r}{r_0}\right)^2\sin (\theta)}{\left[1+\left(\frac{r}{r_0}\right)^2\right]^2}\;,
\end{equation}
where
$\rho_c$ is the the density at the centre of the core, $r_0$ the scale length,
 and $A$ is a factor that determines
how asymmetric is the core.
 This density profile (Fig.~\ref{asym.cores}, left) resembles
the BE sphere density profile (drops as $r^{-2}$ at large
radii and gets flatter near the centre).
In Fig.~\ref{images}, we present our calculations for a core
with $n_c=10^6~{\rm cm}^{-3}$, $r_0=2\times 10^3$~AU,
 $R_{\rm core}=2\times 10^4$~AU, $M=7.3$~M$_{\sun}$ and $e=2.5$ 
 ($e$ is the ratio of the optical depth at $\theta=90$, i.e. looking at an
 edge-on core, to the optical depth at $\theta=0$).

\paragraph{Temperature Profiles}
%%%%%%%%%%%%%%%%%%%%%%%%%%%%%%%%%%%%%%%%%asym.cores
%%%%%%%%%%%%%%%%%%%%%%%%%%%%%%%%%%%%%%%%%%%%%%
The dust temperature is $\theta$ dependent 
(Fig.~\ref{asym.cores}, centre). As expected, the `equator' of the
core is colder than the `poles'. The difference in temperature is
around 3-4 K for the core under study ($e=2.5$). The difference
is expected to be larger for less symmetric cores. 

\paragraph{SEDs and Images}
 The SED distribution 
(Fig.~\ref{asym.cores}, right)
of the core is the same at any viewing 
angle, because the core is optically thin to the radiation it emits
(FIR and longer wavelengths). However, the isophotal maps  
are quite different for different viewing
angles (Fig.~\ref{images}).

At 200 ~$\micron$ the outer, hotter parts of the core dominate
the core emission.  The core appears spherical
when viewed pole-on and elongated when viewed edge-on. The image at
30$^{\rm o}$ is quite interesting, with two absorption `blobs' near the
centre of the core. A quick search through the Kirk (2003)
sample of ISO/ISOPHOT observations did not reveal cores with such 
distinctive features, which is surprising, because one excepts some of
the observed cores to be viewed at such angles.
Further study is required to see whether this is due to the
sensitivity and resolution of observations, due to selection effects,
 or it is connected with the core structure and its environment (i.e. the ambient
 cloud).
%%%%%%%%%%%%%%%%%%%%%%%%%%%%%%%%%%%%%%%% images
%%%%%%%%%%%%%%%%%%%%%%%%%%%%%%%%%%%%%%%%%%%%%%%%

At longer wavelengths, like 850~$\micron$, the core emission is
regulated by the column density. Thus, when the core is viewed edge-on
the intensity is larger at the centre. For the same reason the core looks
elongated when the observer is looking at it at any other
direction than pole-on.

\section{Conclusions}

We applied a Monte Carlo radiative transfer method to spherical cores embedded 
inside molecular clouds. We find that the temperature inside these cores
is less than 12 K, even for an ambient cloud with
moderate visual extinction $\sim 5$.  We also studied 
asymmetric non-embedded cores and the preliminary results
show that a small disk-like asymmetry in the density
distribution will make the core look elongated when viewed at a random
angle.

%------------ end of article ------------------->>
\begin{acknowledgments}
We  acknowledge help from the EC Research Training Network
``The 
Formation and Evolution of Young Stellar Clusters'' (HPRN-CT-2000-00155).
\end{acknowledgments}

\end{document}